\documentclass[prd,twocolumn,amsmath,amsfonts,amssymb,showpacs,nofootinbib]{revtex4}
\usepackage{epsfig,graphicx,bm,color}
\usepackage{url}
\begin{document}

\title{The 130 GeV Fingerprint of Right-handed Neutrino Dark Matter}
\author{Lars Bergstr\"om}
\email{lbe@fysik.su.se}
\affiliation{{T}he Oskar Klein Centre for Cosmoparticle Physics, Department of Physics, Stockholm University, AlbaNova, SE-106 91 Stockholm, Sweden}
%\date{Aug 29, 2012}
%\date{\today}
%\pacs{}

\begin{abstract}
Recently, an interesting indication for a dark matter signal in the 
form of a narrow line, or maybe two lines and/or an internal 
bremsstrahlung feature, has been found in analyses of public 
data from the Fermi-LAT satellite detector. 
As recent analyses have also shown that there is little sign of 
extra contributions to 
continuum photons, it is natural to investigate leptophilic 
interacting massive particle 
(LIMP) models. We show that a model of radiatively generated
 neutrino masses may have the properties needed to explain the 
Fermi-LAT structure around 
130 GeV. This model was proposed some 10 years ago, and predicted a clearly 
observable $\gamma$-ray signal in the Fermi-LAT (then GLAST) detector.
 Here, we update and improve that analysis, and show as an example that a
right-handed neutrino of mass 135 GeV should give rise to 
three conspicuous effects: a broad  internal
 bremsstrahlung bump with maximum around 120 GeV, a 2$\gamma$ 
line around 135 GeV, 
and a $Z\gamma$ line at 119.6 GeV (neglected in the previous work). These features  together
 give a good fit to the 130 GeV structure, given the present energy resolution
of the Fermi-LAT data. An attractive feature of the model is 
that the particle physics properties are essentially fixed, once the relic density 
and the mass of the right-handed neutrino dark matter particle 
have been
 set. Puzzling features of the data at present are a slight displacement of 
the signal from
the galactic center, and a needed boost factor of order $5-15$. This 
presents interesting
challenges for numerical simulations including both baryons and dark 
matter on scales of 100 pc, and perhaps a need to go beyond the 
simplest  halo models.
With upcoming experiments having better energy resolution, or
 with future Fermi-LAT data,
the double-peak structure with a definite predicted 
ratio of the strengths of the two lines and the internal bremsstrahlung feature 
should be seen, if this model is correct. With the planned satellite GAMMA-400,
a striking fingerprint of this dark matter candidate should then appear. 
\end{abstract}
\maketitle

\newcommand{\ga}{\gamma}
\newcommand{\be}{\begin{equation}}
\newcommand{\ee}{\end{equation}}
\newcommand{\bea}{\begin{eqnarray}}
\newcommand{\eea}{\end{eqnarray}}
\newcommand{\ds}{{\sf DarkSUSY}}
\newcommand{\code}[1]{{\tt #1}}

\hyphenation{}

\section{Introduction}

For $\gamma$-rays coming from annihilations of dark matter particles 
in the halo, the Fermi-LAT instrument \cite{fermilat} has very successfully 
delivered bounds that have recently started to probe into the parameter 
space of viable models, 
  in line with pre-launch expectations \cite{prelaunch}, 
in particular for dwarf spheroidal galaxies
\cite{fermidwarfs} and galaxy clusters \cite{fermiclusters}. 

Interestingly, there have very recently appeared analyses of public Fermi-LAT data 
\cite{fermi-pub}  where a rather strong tentative dark matter signal has been seen, 
in a line-search from the halo in the vicinity of the galactic centre 
\cite{bringmann,weniger}  There is a feature, which has been interpreted as 
either originating from internal 
bremsstrahlung \cite{lbe89,IB}
around 150 GeV \cite{bringmann}, or a $\gamma$-ray line \cite{lines} at 
around 130 GeV \cite{weniger}. This discovery has already gained considerable 
attention from theorists (see, e.g., \cite{thpapers}  and \cite{lbereview,bring_wen} for recent reviews). The line signal has subsequently 
been independently verified by other analyses of the same public data 
\cite{tempel,su_fink}. In particular, in the analysis \cite{su_fink} a 
signal of more than $5\sigma$ was claimed, with some evidence also for a second 
line, with energy  consistent with  annihilation into
$Z\gamma$, which generally also should exist in many models \cite{zline}.

There have already appeared quite a number of proposed models
for this feature aound 130 GeV \cite{thpapers}, although the previously
much studied supersymmetric models now seem  disfavoured in most of the supersymmetric scenarios, 
due to the non-appearance of a characteristic $\gamma$-ray continuum that should 
result from fragmentation of quarks, $Z$- and $W$-bosons or 
gluons \cite{lisanti}. (See, however, \cite{bring_wen} for some 
interesting surviving parts of the huge supersymmetric
parameter space.)

Let us now turn to another attractive model from the particle physics
point of view, namely radiative see-saw models for neutrino masses
(for a recent review containing many of the original references, see \cite{seesaw}).

Some 10 years ago, 
an interesting model was proposed \cite{KNT}, where a right-handed neutrino of mass 
less than about a
 TeV plays a crucial role in giving mass to the otherwise massless standard
model neutrinos through a high-order loop mechanism. (This is a version of the
Zee model \cite{zee}. See also \cite{seto04,ma} for a more recent formulation of models with similar phenomenology.)

In this model, neutrino masses appear only at the three loop
level, achieved by supplementing in the simplest version 
the Standard Model (SM) fields with two charged singlet
scalars $S_1$ and $S_2$ and one right handed neutrino $N_R$. (We will here stick to this
simplest model, as we expect the $\gamma$-ray features studied here depend very little
on the details of more elaborate, and perhaps more realistic schemes, 
such as elaborated upon in \cite{seto04,seesaw,review}).  Lepton number is
broken explicitly by including a Majorana mass term for the right-handed
neutrino, and imposing a discrete $Z_2$ symmetry under which the SM fields and
$S_1$ are neutral but $S_2$ and $N_R$ transform as
\begin{equation}
Z_2: \{S_2, N_R\} \longrightarrow \{-S_2, -N_R\} \ ,
\end{equation}
which forbids Dirac masses for the neutrinos.  This gives the Lagrangian
\begin{eqnarray}
{\cal L}& = & f_{\alpha\beta}L_{\alpha}^TCi\tau_2L_{\beta}S_1^{+} +
g_{\alpha}N_RS_2^{+}l_{{\alpha}_R}\nonumber \\
&& + M_RN_R^TCN_R + V(S_1,S_2)
+ {\rm h.c.} \ .
\end{eqnarray}
The potential $V(S_1,S_2)$ contains in particular a $(S_1S_2^{*})^2$ coupling, 
and a  mild hierarchy of masses $M_R<M_{S_1}<M_{S_2} \propto $ 100 GeV - 1 TeV is assumed.
Furthermore, the Yukawa couplings $f_{\alpha\beta}$, $g_{\alpha}$ are of order unity, making
$N_R$ stable in view of the discrete symmetry.  Left-handed neutrino
masses are first induced at three loops \cite{KNT,seesaw}.  
For $M_{S_2}$ of the order of a TeV, one finds an
effective dimension-five effective mass scale of $\Lambda > 10^9$ GeV, giving
remarkably neutrino masses at the $0.1$ eV scale without involving 
fundamental mass scales larger than a TeV. 

In this leptonically interacting massive particle model (a leptophilic WIMP, 
named LIMP in the paper by Baltz and Bergstr\"om \cite{BB}, BB in the following), 
$N_R$ becomes stable and is therefore a 
natural dark matter (DM)
candidate. Through $S_2$ exchange it couples extremely
weakly today, due to the low DM velocities in the galactic halo. The reason for this is its 
Majorana nature, which means that the couplings in the S wave are proportional to the mass of 
charged leptons in the final state \cite{goldberg}. In the early universe, on the other hand, 
P-wave annihilation was important and 
set the relic density $\Omega_{DM}h^2\sim 0.11$ (the scaled Hubble constant 
$h\sim 0.7$). It was noted in the original proposal \cite{KNT} that this would 
be a fine DM candidate, 
but thought to be essentially undetectable, as the rate of direct detection through 
scattering on nucleons would be very small. Similarly, indirect detection through 
neutrinos from the Earth or the Sun will
not be possible, as the cross section for capture is negligibly small.

However, in BB \cite{BB}, it was pointed out that this candidate, on the contrary,
has excellent detection probability in indirect detection through the 
annihilation $N_RN_R\to 2\gamma$ or through the internal bremsstrahlung process
$N_RN_R\to l^+l^-\gamma$ (the last process with its surprising avoidance of helicity suppression, 
and the relation to the 
$2\gamma$ line was first discovered in \cite{lbe89} in the context of 
supersymmetric
models for dark matter, with  refinements over the last few years
\cite{IB,bringmann,bell,urbano}). Thus, this model has the intriguing property of connecting
the dark matter problem with that of neutrinos masses. The unique window is
through $\gamma$-ray detection of annihilation
of the typical fingerprint of this model: an internal bremsstrahlung broad
enhancement with maximum near 90 \% of the $N_R$ mass, supplemented with two
$\gamma$-rays lines from the $\gamma\gamma$ and $Z\gamma$ final states, where
the $\gamma\gamma$ line will appear at $E_\gamma = m_R$ and the $Z\gamma$ line
at $E_\gamma=m_R(1-m_Z^2/(4m_R^2))$, as dictated by energy and momentum conservation.
Both lines are intrinsically very narrow, meaning that in practice the energy
resolution of the detector will be decisive for their detection.
(Note that for Majorana particles, the annihilation in the S-wave takes place
from a pseudoscalar state, and therefore no $H\gamma$ line, with $H$ the 
Higgs boson, is expected, as a $0\to 0$ radiative transition is forbidden.)

In BB,  $Z\gamma$ electroweak mixing and 
therefore the $Z\gamma$ annihilation channel was neglected 
(and in \cite{lbe89} it was not kinematically allowed). 
That this model has a 
very good chance of giving an observable structure in Fermi-LAT is in fact 
strengthened by a more complete analysis, which will be briefly 
presented here. We will find a striking agreement of the shape of the structure
with  the (admittedly still scarce) Fermi-LAT data 
analyzed in \cite{weniger}.\footnote{There are many details of 
the model that are worth studying 
more closely, and that may move the results up or down by a factor of 
two or so (see for example \cite{seto04} where a second righthanded neutrino is shown to be needed) - a small uncertainty compared to many astrophysical unknowns, 
such as the DM density distribution near the galactic centre.} 
A very interesting aspect, which contrasts with  supersymmetric DM,
is the essential lack of freedom of choice of parameters and therefore 
the predictive power of the model.  

The cross determining the relic $N_R$ density can be written \cite{goldberg,BB}
\begin{eqnarray}
\sigma v\left(N_RN_R\to \ell^+\ell^-\right)=
{g_\ell^4\over 32\pi m^4_R(1+f^2)^2} & &\nonumber \\ 
\left[m^2_\ell+{2\over3}\left(\frac{1+f^4}{(1+f^2)^2}\right]\,m_R^2v^2
+\ldots\right],
\label{eq:freeze}
\end{eqnarray}
with $m_\ell$ the charged lepton mass, $m_S=fm_R$ the $S_2$ mass and $m_R$
the LIMP ($N_R$) mass.  For $f$ not too much larger than unity, the
factor in square brackets is typically close to $0.5$, so that for
$m_R$ not larger than a TeV, the P-wave cross section (the term proportional to
$v^2$ in Eq.~(\ref{eq:freeze})) determines the freeze--out temperature
$T_f$ and thus the relic abundance. Typically,
 $T_f/m_R\sim 1/20$, and thus $\langle v^2\rangle_f=6T_f/m_R\sim 0.3$.  The
requirement that the LIMP be the dark matter particle, with relic abundance
$\Omega_R h^2\approx 0.11$, then fixes the freeze-out cross section
\begin{equation}
\sum_{\ell=e,\mu,\tau}
\langle\sigma v\rangle_f\approx {\sum_\ell g^4_\ell(1+f^4)\over 290\pi m_R^2
(1+f^2)^2}\approx 3\times 10^{-26}\ {\rm cm}^3\;{\rm s}^{-1}.
\end{equation}
For a given $N_R$ of mass $m_R$, this sets the normalization of the
combination $\sum_\ell g^4_\ell(1+f^4)/(m_R^4(1+f^2)^2)$ which governs
many annihilation processes. The total annihilation rate into
lepton pairs $\ell^+\ell^-$ at rest (putting $v=0$ in
Eq.~(\ref{eq:freeze})) is for each flavour 
(assuming flavour universal couplings)
\begin{equation}
(\sigma v)_{v\sim 0}\approx
10^{-25}\;\left(\frac{m^2_\ell}{m^2_R}\right)\ {\rm cm}^3\;{\rm s}^{-1}\;.
\label{eq:rest}
\end{equation}

It was noted \cite{KNT,BB} that in order that the couplings $g_\ell$ not be
larger than 1, for nearly degenerate $N_R$ and $S_2$, $m_R$ cannot be too
much heavier than 1 TeV, but could be significantly lighter. 
We now thus insert the value found in \cite{weniger} of around $130$ GeV for 
$m_R$. Actually with $m_R=135$ GeV we find a slightly better fit, and we will
use that value in the following.\footnote{One should be aware that there is an up to a few GeV uncertainty of
the exact value, depending on the present low statistics and 
also factors like the calibration of the Fermi-LAT instrument.
The relative location of the three $\gamma$-ray features is however fixed once the mass is set.} 
Assuming near degeneracy, but choosing for simplicity 
$f=1.1$, to avoid complications of possible coannihilations (this could of 
course be relaxed in a more refined treatment), we find a value of 
$g\sim 0.52$, where we for simplicity assume flavor universality, i.e.,
$g_e=g_\mu=g_\tau=g$ (see \cite{seto04} for a more realistic ansatz in a more elaborate model).  
 LIMPs with these values of the parameters 
have a very small
annihilation rate for the $e^+e^-$ and $\mu^+\mu^-$ channels. Even for
$\tau^+\tau^-$, the helicity suppression is of the order of $10^{-4}$. At that
level, the P-wave term in the galactic halo also may contribute a small amount. 
We neglect 
that here, but include continuum $\gamma$-rays from decays of directly 
produced $\tau$ leptons in our Figures. 
They turn out to be  not very important, however.
Also, there is a contribution to $\gamma$-rays from $Z$ decays produced in the
similar 'gaugestrahlung' process with  the final state $\ell^+\ell^-Z$ which also is
surprisingly large \cite{bell,urbano}. (There is no corresponding 
$W$-strahlung for these singlet fields.) This contribution from $Z$ decays is also included in the numerical estimates, but also turn out to be inessential.

In fact, predictions for a 100 GeV LIMP were made in BB
(see Fig. 2 (a) in \cite{BB}), and shown to be possibly quite spectacular.
(This is in contrast with the predictions for positrons from the same process
which need a ``boost factor'' of more than several thousand to match the
 anomalous positron ratio of which there were indications already at that time \cite{BB}.) 
Thus there is a reason for scrutinizing present Fermi-LAT data to see whether 
the 135 GeV LIMP fits the spectral feature recently 
discovered.\footnote{This feature often is referred to as the ``130 GeV line''. However, as we 
will see, in the present model 
it is a rather complicated composite structure, which is why we prefer 
to call it the ``130 GeV fingerprint'', which for $m_R=135$ GeV consists of a 
135 GeV $\gamma\gamma$ line, a 119.6 GeV $Z\gamma$ 
line and an internal bremsstrahlung continuum, from $\ell^+\ell^-\gamma$.}
 Indeed, we will find that this model, especially when 
augmented with the $Z\gamma$ line (which was left out in BB) and having both that line, 
the $2\gamma$ line and the internal bremsstrahlung contribution fits the Fermi-LAT feature
surprisingly well. We will compare with the original analysis in \cite{weniger}
and choose for convenience the Einasto profile 
Reg. 3 ``SOURCE'' events 
in the parlance of that reference.\footnote{This is the preferred density 
profile, see Fig.~3 in a remarkable recent analysis \cite{bring_wen} -- 
actually, the first to use this tentative signal to constrain details 
of the halo distribution.} 

We note that we are not making a ``best fit'' analysis as this
is not very meaningful given the presently rather scarce data. 
The point is that with more and better data in the future, the 
$\gamma$-ray ``fingerprint'' of this model should
present itself, if this model for the dark matter is correct.

To make sure that continuum $\gamma$s are not overproduced in $Z$ decays 
from the $Z\gamma$ final state, we include this contribution using the 
freely available {\sc DarkSUSY} code \cite{ds}, but also this flux is 
quite small.
In recent versions of the {\sc DarkSUSY} code all three processes: internal 
bremsstrahlung, $2\gamma$ and $Z\gamma$ 
are included and well documented in the original literature 
\cite{lbe89,bbe,bringmann}, \cite{BU,gondolo} and \cite{UB}, 
respectively, so here we 
only show and discuss the results. We note that all three processes
have to exist in this type of model, and have to be included for consistency.\footnote{It would be 
important if a new independent calculation of the 
loop-induced $Z\gamma$ rate could be made, as the results
of the two existing calculations \cite{BU,fawsi} agree 
well in models where intermediate $W$ bosons
dominate, but in the ``bino-like'' case of relevance here, 
there seem to be some discrepancies \cite{fawsi}.} 

\begin{figure}[htb]
\begin{center}
\epsfig{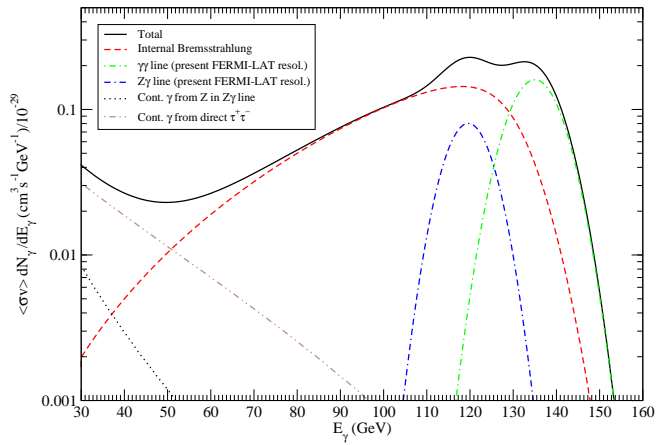}
\end{center}
\caption{The differential photon spectrum for the process $N_RN_R\to
\ell^+\ell^-\gamma$, $\gamma\gamma$ and $Z\gamma$, smeared with the 
present Fermi-LAT energy resolution, $\Delta E/E \sim 0.1$. The total
spectrum is given by the black solid line, the internal bremsstrahlung
by the dashed red line, the smeared $\gamma\gamma$ line by the green
dash-dotted line and the smeared $Z\gamma$ line by the blue
double-dash-dotted line. At the lower left corner the small contributions
from the $\tau^+\tau^-$ final state as well as that from $Z$ decays
can se seen.}
\label{fig:fig1}
\end{figure}

Normalizing to the correct relic density, the internal bremsstrahlung 
contributes around 
$6.7\cdot 10^{-29}$ cm$^3$s$^{-1}$ to the annihilation of 
these slow LIMPs in the halo. The bulk forms a broad bump 
around $E_\gamma\sim 0.9 m_R$ (see Fig.~1).
The $2\gamma$ line gives $2.3\cdot 10^{-29}$ cm$^3$s$^{-1}$ 
(including a factor
2 for the two photons),
and the $Z\gamma$ line has a $(\sigma v)_{v\sim 0}$
of $1.0\cdot 10^{-29}$ cm$^3$s$^{-1}$. The combined rate in the 
galactic halo is thus
\begin{equation} (\sigma v)^{tot}_{v\sim 0}= 1.0\cdot 10^{-28}\ {\rm cm}^3{\rm s}^{-1},
\end{equation}
a value not to far from the  estimate of the experimentally observed
effect in \cite{weniger} of 
$(\sim 12.7\pm 5.7^{+3.2}_{-2.8})\cdot 10^{-28} $cm$^3$s$^{-1}$ 
(for the favoured Einasto  profile).
In \cite{weniger}, this was interpreted only in terms of one $\gamma$ line, 
but the limited
energy resolution of the Fermi-LAT instrument, around 
$10$ \% FWHM \cite{fermiline}, 
is only barely enough to resolve the structure into its 
separate components (see Fig.~\ref{fig:fig1}).
Note, however, that \cite{su_fink} favour, although not very strongly, 
a two-line 
structure, especially when selecting data in the detector from 
directions where the photons 
pass more matter in the calorimeter and the energy resolution is 
therefore improved by a factor around two
\cite{su_fink}. We will soon return to our prediction for an 
instrument with this better energy resolution.

In Fig.~\ref{fig:fig1} is shown the total $\gamma$-ray differential flux
and the respective contributing process. The results are given
for the particle physics factor $d(N_\gamma\sigma v)/dE_\gamma$ for slow 
annihilations such as in the galactic halo,
i.e., in units of cm$^3$s$^{-1}$GeV$^{-1}$, and are very 
robust around this mass range, as the
normalization is set by the relic density constraint. 

\begin{figure}[htb]
\begin{center}
\epsfig{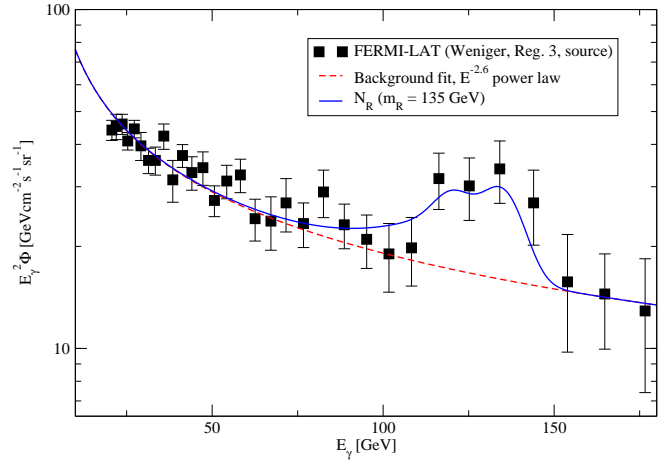}
\end{center}
\caption{Comparison of the total $\gamma$-ray differential energy 
results (multiplied by $E^2$) for a 
135 GeV right-handed neutrino dark matter
candidate with the Fermi-LAT public data \cite{fermi-pub}, 
as analyzed by Weniger \cite{weniger}.
A simple power-law fit $\sim E^{-2.6}$ to the continuous background 
has been made, and is also shown.}
\label{fig:fig2}
\end{figure}

In Fig.~\ref{fig:fig2} a comparison with of these results is made 
with the Fermi-LAT data as analyzed by Weniger \cite{weniger}, 
multiplied by $E^2_\gamma$, and and arbitrary normalization constant
(of order 10) to match the overall size of the effect. 
The reason for this ``boost factor'' 
is presently unknown, but in this model it has to be explained by 
astrophysical effects, such as the detailed distribution of dark matter 
near the galactic centre. (So-called Sommerfeld enhancement \cite{hisano1}
is not expected in this model. There may in principle be 
fine-tuned mechanisms like 
S-wave pseudoscalar resonances or particles with higher electric charge 
running the loop, but we do not employ such exotica here.) 
The required boost may be related to another puzzle
of the signal, which is a displacement from the exact galactic 
centre by around 200 pc.\footnote{In fact, a preprint recently appeared
\cite{kuhlen} where such a displacement is shown not to be unnatural 
in simulations of the combined baryon and dark matter system. It remains
to be seen whether the larger than expected density near the emission region
can also be explained by similar effects. The off-set could perhaps also be explained
by the low statistics of the tentative signal \cite{fan}.}  

\begin{figure}[htb]
\begin{center}
\epsfig{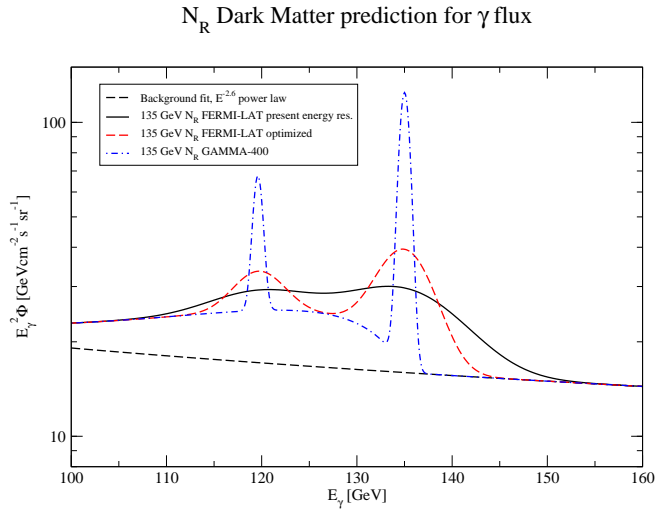}
\end{center}
\caption{The $\gamma$-ray differential energy results (multiplied by 
$E^2$) for a 
135 GeV right-handed neutrino dark matter
candidate are shown,  with the present Fermi-LAT energy resolution 
$\Delta E/E=10$ \% FWHM (solid black line), with a factor of 2 improvement
(red dashed line) and with a future $\gamma$-ray instrument, 
such as GAMMA-400 \cite{gamma-400} 
(dash-dotted blue line) with resolution at the one percent level. 
The extrapolated power-law $\sim E^{-2.6}$ of the presently measured continuous 
$\gamma$-ray background is also shown.}
\label{fig:fig3}
\end{figure}

As can be seen, once the overall strength has been set, the agreement 
with present data is (perhaps fortuitously) intriguing. For the 
rather low average energy resolution (10\% FWHM) of Fermi-LAT, the double-peak
structure is barely visible. Improving, however, the resolution by a factor
of two, which has been 
done in \cite{su_fink} (sacrificing statistics), by selecting observation angles
which give large path-lengths of the electromagnetic shower in the detector,
the twin peak structure is much clearer. This is qualitatively in 
agreement with our results
(see Fig.~\ref{fig:fig3}).\footnote{It
has been suggested in \cite{su_fink} that Fermi-LAT may change its observational
search strategy so as to favour these side-ways entering events. If this can be
done technically, this interesting proposal could mean that Fermi-LAT
may establish the existence of this dark matter fingerprint with high confidence
over the next couple of years.}

In Fig~\ref{fig:fig3} is also shown what one may expect from the next generation
of $\gamma$-ray space detectors with energy resolution at the one percent level,
such as GAMMA-400 \cite{gamma-400} and DAMPE (see \cite{dampe} and 
references therein). 
Given that the type of model
described here is the correct explanation of dark matter, 
the features of the signal
would be striking. With such an instrument one could start 
analyzing the dark matter
halo density distribution in some detail. In fact, the property 
of the fingerprint 
of this model is, besides the two strong lines, the rather broad and slightly
asymmetric bremsstrahlung bump. The absence of this bump 
would rule out the model.

The theoretical reason for the necessity of the internal 
bremstrahlung bump and its relation
to the line signal is quite interesting. It was shown in 
\cite{lbe89} how these features
are crucial for reproducing the effective axial anomaly in
 these theories (which lack
anomalies at the fundamental level). In fact, the strength 
of the $\gamma\gamma$ line
can almost trivially be computed by using the anomaly result 
($|F|=1$ in \cite{lbe89}). Also, compact formulas for the 
internal bremsstrahlung contribution can be found there
(recently checked independently \cite{urbano}).
The validity of these formulas is more general than for the specific 
 $N_R$ case discussed here.
The strength of the $Z\gamma$ line is trickier to compute due to 
the non-zero $m_Z$, but we can 
use \ds\ (based on \cite{UB}) for its calculation.

Of course, one should also be aware
of the possibility that the feature in the Fermi-LAT may be spurious. 
Although it seems  
unlikely that it is caused by a statistical fluctuation, one cannot 
yet exclude that its origin is in some (unknown) instrumental effect. Hence 
one should wait for the results of the analysis of the Fermi-LAT 
collaboration itself 
before taking our comparison with data too seriously.

It will also be interesting to see if data can 
be independently reproduced by the HESS-II instrument which just has 
seen first light (see \cite{farnier},
where also the potential of CTA \cite{cta} and an instrument like GAMMA-400 
\cite{gamma-400} is analyzed).  Also, it remains to be seen 
if the recent rather statistically shaky indications of a signal 
from galaxy clusters \cite{hektor2}
will survive more data. This will  most immediately come from the Fermi-LAT 
instrument, which will continue to collect an amazing amount of 
impressively clean and interesting data over the next few years.

The type of model discussed here has been rather extensively studied in 
the neutrino and particle physics communities in recent years, 
thanks to their ability to explain the observed neutrino masses without 
invoking a super-high mass scale. If couplings are not flavour diagonal,
dangerous non-observed decays like $\mu\to e\gamma$ may occur. However,
this can be rather easily avoided, without affecting appreciably the results
presented here, by invoking discrete symmetries, as 
explained in \cite{seesaw}. In that reference, it is also pointed out that,
remarkably, in the case of having $m_R$ and $m_S$ of the order of 100 GeV
and $g \sim 0.5$, i.e., precisely the range of  values that are 
forced upon us if we want to explain the Fermi-LAT structure, 
also the
enigmatic discrepancy of the measured $(g-2)_\mu$ value \cite{g-2} with the 
theoretical prediction \cite{marciano} has a chance to be explained. 
(See, however, \cite{bringmann} for a problem in a related generic model.)
This is an interesting topic for further studies.

Of course, even if the present indications from the public Fermi-LAT 
would disappear, this model still gives an 
interesting signal  to search for using
present and new $\gamma$-ray detectors. The signal would naturally
appear between 100 GeV and a TeV (perhaps, then, with its canonically 
predicted rate, a factor of 10 lower
than that indicated in \cite{bringmann,weniger}). This energy 
region will be closely watched in the near future \cite{farnier}.

One should note that there are interesting very recent indications that the dark
matter density, at least locally, may be a factor of around 3 higher
than the generally adopted value \cite{read}. It would be interesting to
analyze the predicted rates in such models with either a dark disk and/or
an oblate dark matter halo. This could in principle explain the boost of order
10 needed to fit the $N_R$ dark matter model to the Fermi-LAT 
data as analyzed by Weniger \cite{weniger}.  

Finally, a note on other ways to test the hypothesis of LIMP dark matter.
Unfortunately, the scattering cross section on nuclei will be very small. This
was recently shown \cite{effective} for an effective description of the 
$2\gamma$ vertex. Also the radiative scattering process with one photon
exchange responsible for scattering on nuclei has a depressingly low rate
(see \cite{aminne} for a detailed explanation of the reason for this 
suppression). 

The low scattering rate also 
means a very low capture rate, which hurts detection by neutrinos from the 
Sun or Earth.

Whether there will be traces of this new sector, only interacting leptonically
with standard model particles, at the LHC remains to be seen. In principle,
there could be mediators that for instance also would explain the somewhat 
too high $2\gamma$ decay rate of the Higgs boson candidate indicated by recent
LHC data \cite{lhc}. It would then, similarly to in our case, be important
to detect and measure the rate of $H\to Z\gamma$ (\cite{lbe85,djouadi,gainer}), 
which in the standard model is closely related
to $H\to 2\gamma$.

Another possibility for a future linear $e^+e^-$ collider may be 
contemplated.  The $S_1$ and $S_2$ states have unit electric charge and 
would appear as (rather slowly rising) enhancements in the total $e^+e^-$ cross section, both 
starting around 150 GeV  energy per beam.

There is actually another, challenging, idea to directly probe the dark matter
nature of the right handed neutrino. The idea is to excite with a 
strong electron beam ambient $N_R$ LIMPs to the nearly degenerate $S_2$, which
would then immediately decay isotropically back to an electron 
and $N_R$. This was first suggested and computed in the supersymmetric 
scenario in \cite{aminne}, and recently rediscovered \cite{hisano}. 
It seems that substantial experimental development would be needed, however.

It would also be interesting to work out the possible $\gamma$-ray 
signals that could appear
from very powerful electron jets emanating from centres of active galaxies, 
along the ideas of Ref.~\cite{gorch}, for a 135 GeV $N_R$. 

Multiwavelength studies of the region around the galactic centre,
in particular radio data, could be important for these models with positron and
electron emission (see, e.g.,\cite{regis}). It seems, judging from a recent 
detailed study, however, that for the Einasto profile current limits are 
not constraining \cite{laha}, but maybe with future experiments this 
could give an important cross-check.

To conclude, for indirect detection of dark matter, the excellent 
Fermi-LAT data -- made public to the scientific community with many excellent 
analysis tools -- has opened a very exciting opportunity for studies of 
detailed predictions from models
such as the one discussed here. Within a few years, the aquired 
data (and here the verdict of the Fermi-LAT collaboration itself
will be important) could make us 
accept the model with confidence -- or disprove it. 
These are interesting times for dark matter studies, indeed.   

%%%%%%%%%%%%%%%%%%%%%%%%%%%%%%%%%%%%%%%%
\smallskip
\acknowledgments
The author wishes to thank Ted Baltz for the collaboration on the early 
BB paper, and Gianfranco Bertone, Torsten Bringmann,  
Jan Conrad, Joakim Edsj\"o, Christian Farnier and Christoph Weniger 
for recent useful discussions.
The research of the author was carried out under Swedish Research Council (VR) 
contract no. 621-2009-3915.
%%%%%%%%%%%%%%%%%%%%%%%%%%%%%%%%%%%%%%%%%%%%%%%%%%%%%%%%%%%%%%%

\end{document}